\documentclass[preprint]{aastex}

\newcommand\mdot{\rm \dot{M}}
\newcommand\msun{\rm M_{\odot}}
\newcommand\lsun{\rm L_{\odot}}
\newcommand\rsun{\rm R_{\odot}}
\newcommand\msunyr{\rm M_{\odot}\,yr^{-1}}
\newcommand\be{\begin{equation}}
\newcommand\en{\end{equation}}

\newcommand\kms{\rm{\, km \, s^{-1}}}

\newcommand\etal{{\rm et al}.\ }

\begin{document}

\title{Comments on Inferences of Star Formation Histories
and Birthlines}

\author{Lee Hartmann}
\begin{abstract}
Palla \& Stahler have recently argued that star formation in Taurus
and other nearby molecular clouds extends over a period of at least
10 Myr, implying quasi-static cloud evolution and star formation.  
Their conclusions contradict
other recent results indicating that molecular clouds
are transient objects and star formation proceeds rapidly.  
The Palla \& Stahler picture implies that most molecular clouds
should have extremely low rates of star formation, and that in such
inactive stages the stellar initial mass function should be strongly
skewed toward producing stars with masses $\gtrsim 1 \msun$; neither
prediction is supported by observations.
I show that the Palla \& Stahler conclusions for Taurus 
depend almost entirely on a small number of stars 
with masses $\gtrsim 1 \msun$; the lower-mass stars
show no evidence for such an extended period of star formation.
I further show that most of the stars apparently older than 10 Myr in 
the direction of Taurus are probably foreground non-members. 
I also present birthline calculations which support the idea that 
the ages of the stars with masses $\gtrsim 1 \msun$ 
have been systematically overestimated because
``birthline'' age corrections have been underestimated; 
such birthlines would eliminate the need to
postulate skewed initial mass functions.  The simplest
and most robust explanation of current observations characterizing the
vast majority of young stars in molecular clouds is that cloud
and star formation is rapid and dynamic.
\end{abstract}
\affil{Harvard-Smithsonian Center for Astrophysics, 60 Garden
St., Cambridge, MA 02138;\\
Electronic mail: hartmann@cfa.harvard.edu}


\section{Introduction}

Several authors have recently suggested that the old quasi-static
models of low-mass star formation (e.g., Shu, Adams, \& Lizano 1987)
are inadequate, and that star formation is a rapid and dynamic process
(Ballesteros-Paredes, Hartmann, \& Vazquez-Semadeni 1999 $=$BHV99; Elmegreen 2000;
Pringle, Allen, \& Lubow 2001; Hartmann, Bergin, \& Ballesteros-Paredes
2001 $=$HBB01).  Key constraints in the observational case for the timescale
of star formation are the ages of the stellar populations in star-forming
regions.  Using the mean ages of populations, HBB
argued that nearby molecular clouds form rapidly, form stars rapidly,
and disperse rapidly (within a few Myr).

In contrast, Palla \& Stahler (2000 $=$ PS00; 2002, $=$ PS02), 
using some of the same stellar data as HBB01, claimed to find 
evidence for recent ``accelerating'' star formation in molecular clouds,
preceded by very extended periods of time with low star formation rates.
Palla \& Stahler argue that the inferred extended periods of low star formation
are consistent with the old models of quasi-static, slow star formation.
PS02 also take issue with several points I made concerning age distributions
in a previous paper (Hartmann 2001a $=$ Paper I).

In the following, I attempt to frame the issues in a clear and systematic
way, focusing on the recent analysis of the Taurus molecular cloud complex
by PS02, but drawing some broader conclusions as well.  I show that much
of the apparent evidence for an extended period of star formation in Taurus is due
to the stars with masses $\gtrsim 1 \msun$, not the lower-mass stars
which dominate the population.  I present evidence that
the stars with ages $\gtrsim 10$~Myr are strongly contaminated by foreground
non-members.  I further argue that the ages of the younger $\gtrsim 1 \msun$ stars 
have been systematically overestimated by underestimating the ``birthline''
correction to their ages; theoretical calculations of birthlines for cold disk
accretion are presented to support this hypothesis.
My conjecture that birthlines need to be revised avoids the necessity 
of assuming that the stellar initial mass function varies strongly with age, 
a position for which there is no clear observational evidence.  I conclude 
that the observations are not consistent with the picture presented by PS00 
and PS02, and the star formation is rapid and dynamic, as illustrated by the 
age distribution of the great majority of the low-mass stars.

\section{Implications of the PS00/02 picture}

Figure 1 shows the stellar age distribution in Taurus-Auriga inferred by PS02,
derived from data downloaded from their website 
(http://www.arcetri.astro.it/$\sim$palla/taurus/table).
I have binned the stellar distribution in units of 2 Myr 
in order to display some of the apparently older stars; these
older objects are in the table of PS02, but are not shown in the PS00/PS02 papers, 
in which the data are binned in 1 Myr increments from 0 to 10 Myr.
Note that I have changed the age of HBC 412 (Herbig \& Bell 1988 catalog)
from the 20.4 Myr to 4 Myr
(see Table 1); the former apparently was a typographic error in the PS02 table, given
the reported effective temperature and luminosity.  

Figure 1 seems to indicate
a recent rise or burst of star formation, preceded by a very long period of
relatively little activity, with no statistically significant additional peak
in the star formation rate.
The distributions of stellar ages in other regions found by PS00 are qualitatively
similar to that shown in Figure 1.  The eight star-forming regions with molecular gas 
studied in PS00
-- Taurus, Lupus, Chamaeleon I, II, $\rho$ Ophiuchus, IC 348, the Orion Nebula
Cluster, and NGC 2264 -- all show the same behavior.  Of these eight regions,
PS00 found that six of them exhibit peaks in their star formation rates over 
the last 1 Myr; the other two regions peaked 1-2 Myr ago.  
Even Upper Scorpius, which has no significant associated molecular gas, 
exhibited a strong peak in its star formation rate only 2-3 Myr ago.

If these results are taken at face value, one must ask: what is so special about
the last 1-2 Myr?  Why should star forming regions separated by hundreds of pc
(NGC 2264 is at a distance of about 800 pc) form most of their stars at the same time,
if their overall lifetimes are $\sim 10$~Myr?
More broadly, Figures 1-5 of PS00, like Figure 1 of this paper, appear to
show that all the star forming regions have spent most of the last 10 Myr forming
stars at an extremely low rate.  So where are these inactive regions?  The clear
implication of the star formation histories found by PS00 is that more than half
of all molecular cloud complexes should be forming very few stars, much lower than
the typical efficiencies seen in star-forming regions.
HBB01 considered studies of nearby clouds and found
only one plausible example out of nine of low or absent star formation, the Coalsack; 
PS02 point out another nearby cloud, Cha III, which appears to harbor little star formation.
But, according to the PS00/PS02 scenario, {\em most} star-forming clouds should be
inactive, not the small fraction found so far.

Conversely, suppose one adopts the view of HBB01, Elmegreen 2000, and Pringle \etal (2001),
in which molecular cloud and star formation episodes are concurrent, and last only a very few
Myr.  Then there is no need to explain why all the molecular cloud regions peaked
in their star formation rates over the last 1-2 Myr; that is simply the characteristic
timescale for star and cloud formation.  In the rapid formation picture, 
star formation events are not coordinated over hundreds of pc; the present epoch is 
not special in any way.

The basic implication of the PS00/02 scenario is that most nearby molecular clouds
of substantial mass ($10^4 - 10^5 \msun$, i.e., from Taurus-like clouds to those
similar to Orion) should spend most of their lives in a very inactive state of star formation,
with no obvious increase in star formation rates toward the present
(``acceleration'' in the language of PS00/PS02).
As current observations contradict this prediction, the PS00/02 interpretation is suspect.

A clue to the nature of the problem is given in Figure 1, where I have singled
out stars with effective temperatures $\geq 4350$~K, corresponding to stars
of spectral types $\sim$~K5 and earlier (and roughly at masses of
$1 \msun$ and larger for stars on convective tracks,
according to the PS02 tracks).  As shown in Figure 1, these hotter and generally
more massive stars show a distinctly different apparent age distribution than
that exhibited by the cooler, lower-mass stars, and 
dominate the extended tail of the distribution to larger ages.  
Again, taking the PS02 results at face value, one would have to conclude
that the hotter, higher-mass stars formed with a qualitatively different
age distribution than the lower-mass stars in Taurus; the hotter stars formed
first, with little apparent acceleration to the present.  This seems very unlikely; 
why should $\gtrsim 1 \msun$ stars form so differently in time than $< 1 \msun$ stars? 

In the next two sections I consider two factors which could reduce or eliminate the
discrepancy between the age distributions of low- and higher-mass stars,
following the discussion of Paper I; contamination by non-members and errors in isochrones.

\section{Membership and Li}

Associating classical T Tauri stars (CTTS) with their natal molecular clouds is usually
straightforward; these objects, actively accreting from circumstellar disks, and exhibiting
strong line and continuum excess emission, are distinctive and relatively rare.
Identifying weak-emission T Tauri stars (WTTS) which belong to a cloud is much more
difficult, because the most easily-applied indicators -- chromospheric H$\alpha$ emission 
and coronal X-ray emission -- do not decay rapidly with age between 1-100 Myr in
low-mass stars (Brice\~no \etal 1997).  As star formation has continued in the solar
neighborhood for the last 100 Myr (for example, the Pleaides, $\alpha$ Per, and IC 2602/2391 clusters;
Stauffer \etal 1997), these older stars will form a dispersed population, some of which
will lie in front of Taurus.  Since parallaxes are not available for most of these stars
(e.g., Bertout, Robichon, \& Arenou 1999), {\em assuming} that these foreground objects
are members of Taurus, and therefore assigning the cloud distance to them
(140 pc; Kenyon \etal 1994), will result in an overestimate of the stellar luminosity
and an underestimate of the stellar age.  In turn, an underestimated age will make
it appear more likely that such a star is a real Taurus member.  Given the relatively
flat age distribution of the hotter stars in Figure 1, one must consider the possibility
of foreground contamination seriously.

Detection of Li I 6707 \AA\ absorption has come to play a crucial role in identifying
WTTS.  Li can be depleted due to fusion at central 
stellar temperatures $\sim 3 \times 10^6$~K; 
however, the timescale over which this depletion occurs exhibits
considerable sensitivity to stellar mass. Stars with masses $\sim 0.5 \msun$
remain completely convective for most or all of their pre-main sequence
evolution; evolutionary models predict that depletion should begin at
ages $\sim 10-20$~Myr, and proceed rapidly.  On the other hand, stars with
masses $\gtrsim 0.9 \msun$ develop radiative cores, which limits mixing
to the core and thus strongly retards the depletion of Li.  Thus, the utility
of Li as an indicator of pre-main sequence stars, and the age limit it
provides, depends strongly on the mass in question.

Figure 2 shows Li equivalent widths as a function of effective
temperature for differing samples.  The solid dots are young T Tauri stars, while
the solid segmented curve denotes the upper envelope of Li equivalent widths 
for Pleiades members, which have an age of roughly 100 Myr.  
While there is strong depletion at low effective temperatures
in the pre-main sequence phase, there is much less pre-main sequence depletion in the hotter
stars, for the reasons noted in the preceding paragraph.  The dotted segmented
curve shows the upper envelope for stars in the IC 2391/2602 open clusters, which
are estimated to have ages of $\sim 30-50$~Myr (see discussion in Randich \etal 1997).

Figure 2 shows that the detection of Li in a star cooler than about 4200 K is
a clear indication of pre-main sequence status.  However, detection of Li at higher
effective temperatures is not enough; the equivalent width must be larger than that of the
upper envelope of IC 2391/2602 to make it probable that a star is younger than
30-50 Myr.  Failure to recognize this led to the claimed discovery of large numbers
of new WTTS associated with Taurus and other regions (e.g., Wichmann \etal 1996);
more careful followup showed that a large fraction of these stars had Li abundances
consistent with Pleiades age, if not older (Brice\~no \etal 1997; Neuh\"auser \etal 1997;
Wichmann \etal 2000).  

Figure 2 also shows Li equivalent widths, where available, for the apparently older,
hotter Taurus stars (also summarized in Table 1).  There are 11 stars with ages $>$ 10 Myr;
of these, 8 have Li measurements available.  Of these 8, 5 stars have Li equivalent
widths at or below the IC cluster line (RXJ0405.7, RXJ0406.7, HBC 388,
HBC 392, RXJ0441.8), suggesting that they are
$\gtrsim 30-50$ Myr old and thus are unlikely members of Taurus.  These stars are
hatched in Figure 1. One other star (RXJ0423.7) is just slightly above the IC cluster
line.  Thus, at least half, and possibly as many as $\sim$~80\%, of the population
with apparent ages of 10-25 Myr in Figure 1 probably have ages in excess of 30 Myr
and thus are likely older, foreground, stars which are not members of Taurus.
Furthermore, as many of these objects are probably foreground systems, their true
ages are should be older than indicated in Figure 1 because of the overestimate
in distance (e.g, Brice\~no \etal 1997).  These conclusions are consistent with
deep imaging which has provided no evidence for a population of older pre-main sequence
objects among the lower-mass stars (Brice\~no \etal 1998, 1999); main-sequence low-mass
stars will be much fainter and thus drop out of current surveys.

Sometimes association with a cloud is argued on the basis of radial velocities or
proper motions.  Motions alone do not provide a reliable criterion,
as stars younger than about 100 Myr
exhibit relatively small velocity dispersions about the local standard of rest.
For example, consider several stars studied by Walter \etal (1988) that exhibit
Li equivalent widths below the Pleiades boundary.  The 
stars HBC 352, 354, 355, 372, 388, and 392, and SAO 76411A, have a mean
radial velocity and dispersion of $15.3 \pm 3.5 \kms$; this is near the Taurus mean
velocity, and the dispersion, is not much larger than that of the 
obvious members (Hartmann \etal 1986).

A useful test for contamination would be to have equivalent measurements 
for a field off molecular gas; the difference between cloud and non-cloud
regions should provide some indication of the population not associated
with cloud material.  The study by Neuh\"auser \etal (1997) of 
RASS sources in a region south of Taurus is of interest in this connection,
because it covers a comparable area to Taurus but off the molecular cloud, and
has a comparable X-ray detection limit.
Neuh\"auser \etal (1997) report nine K stars in this southern region
with Li equivalent widths comparable to or larger than those 
of IC 2391/2602; moreover, Neuh\"auser \etal
report that most of these stars have radial velocities consistent with
Taurus.  As this region
extends much further away from the galactic plane than Taurus, it is reasonable
to suppose that these results underestimate the number of relatively young
field stars in the direction of Taurus that are not associated with molecular gas.
These results support the contention
that most of the stars with ages $\gtrsim 10$~Myr in Figure 1 and Table 1 are
not true members of the present Taurus clouds.

It should be emphasized that PS00/02 do not consider stars with apparent ages
larger than 10 Myr in their detailed analysis, even though they are claimed to be
vetted members of Taurus in their table of 153 objects.
This age limitation seems to avoid most of the obvious Taurus membership problems.
However, the situation is less clear for other, more distant clusters 
at lower galactic latitudes studied by
PS00, such as the Orion Nebula Cluster and NGC 2264, where Li measurements
are generally not available, and foreground contamination is much more likely.
The possibility of non-members in samples needs to be considered more carefully
when assessing the PS00 results for these more distant regions.

\section{Isochrones and birthlines}

Restricting attention to Taurus stars with ages $\leq 10$~Myr, as in PS02,
and assuming non-members are not present, a problem still remains with the PS02 results; 
the distribution of ages among the 
stars with $T(eff) \geq 4350$~K is dramatically different from the age distribution
of the cooler stars.  A two-sided Kolmogorov-Smirnov test (Press \etal 1986) on
the unbinned age data shows that the two samples have a probability of only $\sim 1.5 \times 10^{-3}$
of being drawn from the same underlying distribution.  Moreover, this pattern --
in which $\sim 1-3 \msun$ stars appear to be considerably older than the $< 1 \msun$ stars --
is seen repeatedly in many regions, such as the Orion Nebula Cluster
(Hillenbrand 1997) and in Upper Scorpius, IC 348, and NGC 2264 (Hartmann 1999; see PS00 results).  

Thus, the PS00/02 age distributions imply that star forming regions pass through an extended 
phase in time during which $1-2 \msun$ stars were produced at a far greater rate than the
lower mass stars.  As shown in Figure 1, the implication of the PS02 results is that
for the first $\sim 5$~Myr, Taurus formed stars with a ratio of $\gtrsim 1 \msun$ stars
to $< 1 \msun$ stars that was roughly an order of magnitude higher than it is today.
Similar results apply to the other regions studied by PS00.  I am unaware of any
evidence for such extremely non-standard IMFs (see Kroupa 2002 for a review).  
Again, the observations indicate that the PS00/02 results must be treated with
skepticism.  

This raises an obvious question: are the isochrones in error?  And if so,
does the problem lie mostly with the higher-mass or lower-mass stars?
Baraffe \etal (2002) provide a recent discussion of the difficulties involved in calculating
evolutionary tracks.  Here I summarize some basic issues, following along
the lines of the discussion in Paper I and also in Hartmann (2001b).

The age of a pre-main sequence star (the elapsed time after the end of protostellar accretion)
can be approximated by the relation
\be
t ~\simeq ~ \eta G M^2 / (R L) ~-~ \eta_{\circ} G M^2 / (R_{\circ} L_{\circ})\,, \label{eq:t}
\en
where $M$, $R$, and $L$ are the stellar mass, radius, and luminosity, respectively,
$\eta$ is a coefficient of order unity which depends upon the internal structure
of the star (see, e.g., Hartmann 1998; Paper I), and subscript zero indicates quantities
at the end of protostellar accretion.  The initial stellar radius and luminosity $R_{\circ}$
and $L_{\circ}$ correspond to the ``birth'' position of the star, and the loci of such points
for stars of differing mass has been called the ``birthline'' (Stahler 1983, 1988).

Since the stellar luminosity can be determined reasonably well for most
objects if the distance is known, and the radius can be inferred from the luminosity 
and effective temperature with reasonable accuracy, the main uncertainty in the first
term of equation (\ref{eq:t}) has long been the stellar mass.  
In the past, it was necessary to estimate
masses from theoretical evolutionary tracks, which included substantial uncertainties in
stellar photospheric boundary conditions and treatment of convection (see the discussion
in Baraffe \etal 2002).  The mass uncertainty, and thus the age
uncertainty, was greatest for lower-mass stars; the radiative tracks for the higher-mass stars
calculated by different groups with differing assumptions
were in better general agreement.

A recent development of great importance is the direct determination
of T Tauri masses from the measurement of Keplerian rotation in their circumstellar
disks, using mm-wave observations of CO emission (Dutrey, Guilloteau, \& Simon 1994;
Guilloteau \& Dutrey 1998; Simon, Guilloteau, \& Dutrey 2000).  For several Taurus stars,
Simon \etal (2000) find internal errors of less than 10\% in the mass, with the overall
uncertainty dominated by the uncertainty in the distance.  Moreover, 
the quantity $L/M^2$ is independent of distance in the Simon \etal (2000) analysis; 
for several of their program objects, Simon \etal estimate errors of 
$<$~10\% in this quantity.  This means that the remaining error
in the first term of equation (\ref{eq:t}) 
is mostly in the radius $R$, which is directly proportional to the distance.
Typical estimates of this error for Taurus CTTS are on the order of 15\% (Paper I; Simon \etal
2002).  Moreover, although most of the stars studied   
are lower-mass objects, one, LkCa 15, is a K5 star with $T(eff) = 4350$~K.  
Simon \etal (2000) show that the
masses estimated from the evolutionary tracks of PS00/02 (from Palla \& Stahler 1999 $=$ PS99)
and Baraffe \etal (1998) are in good agreement with the observations. 
Thus it seems very unlikely that errors in 
the first term of equation (\ref{eq:t}) can account for the discrepancy in age
distributions shown in Figure 1.  

The other major uncertainty in pre-main sequence ages arises from considering the
initial position in the HR diagram from which pre-main sequence contraction proceeds, 
represented by the second term of equation (\ref{eq:t}).
Stahler (1983, 1988) showed that the process of protostellar accretion was likely
to terminate with the star having a finite, relatively small radius, after which
typical pre-main sequence evolutionary tracks apply.  The birthline limits on
stellar radii and luminosities are in part the result of the very temperature-sensitive 
fusion of deuterium in the stellar interior, which prevents contraction below the 
``deuterium main sequence'' until most of the deuterium has been burned.  The latest
version of these calculations is given by PS99, and is employed in PS00 and PS02.

Figure 3 shows the Taurus HR diagram, using the data in the PS02 table.  
I have superimposed isochrones from Baraffe 
\etal (1998), which are similar to those employed by PS99/00/02.
The dotted curve shows the PS99/00/02 birthline; it lies well above the 1 Myr
isochrone.  Therefore, the PS99/00/02 birthline has little effect
on the inferred ages of most of the stars with masses $\gtrsim 1 \msun$.  

However, the birthline calculations of Hartmann, Cassen, \& Kenyon (1997; HCK97) yielded
different results.  To illustrate this in greater detail,
I have recalculated two birthline tracks of HCK97 for accretion
rates of $10^{-5} \msunyr$ and $2 \times 10^{-6} \msunyr$, assuming the cold accretion limit
($\alpha$ = 0 in the HCK parameterization; see also Siess, Forestini, \& Bertout 1997, 1999).
I have also revised the calculations to use a fit to the Baraffe \etal (1998) evolutionary tracks
for the stellar luminosity as a function of mass and radius, and have extended the calculations
in mass until radiative cores form, at which point the evolution cannot be followed further
using the simple polytropic model of HCK97.  

As shown in Figure 3, the revised ``HCK'' birthline for a protostellar accretion rate of 
$10^{-5} \msunyr$ is somewhat lower than the PS99/00/02 birthline (which also assumes the 
same accretion rate).  However, the birthline for $\mdot = 2 \times 10^{-6} \msunyr$ is markedly lower,
crossing the 1 Myr isochrone at $T(eff) \sim 4000$~K.  (The results differ only slightly
from those shown in HCK97 for the same parameters.)  These tracks imply significantly
larger birthline corrections to the ages in equation (\ref{eq:t}) than implied by the PS99 birthline.  

As already mentioned, the simple polytropic approximation used by HCK97 is limited 
to completely convective stars, and thus cannot be extended directly to higher mass stars 
with radiative cores.  However, the trajectories of the HCK tracks
strongly suggest that a continuation of the calculation to higher effective temperatures 
and masses would result in much lower birthlines, and therefore considerably lower ages,
than in the PS99/00/02 analyses.

The difference between the PS99 and HCK97
calculations appears to lie in the differing boundary conditions assumed
(see the discussion in HCK97).  The PS99 birthlines
have been calculated assuming spherical accretion, which is implausible given
the angular momenta of protostellar cloud cores; much if not most of the
stellar mass must initially fall in to a disk, and then be accreted from
the disk.  The HCK97 calculations explicitly assume that protostellar accretion 
is from the circumstellar disk, either through narrow boundary layers or magnetospheric 
accretion shocks covering small areas, as in T Tauri stars;
the latter assumption is consistent with observations suggesting magnetospheric accretion in
some protostellar objects (Muzerolle, Hartmann, \& Calvet 1998; Folha \& Emerson 1998, 2001).
Thus, in the HCK97 case, accretion affects only a small fraction of the
stellar surface; most of the star is free to radiate into space.  This extra energy loss
in comparison with the PS99 calculations causes the star to fuse deuterium faster to replenish
the extra energy radiated into space; in turn, the
faster reduction of deuterium means that the star contracts below the deuterium main sequence sooner,
i.e. at a lower mass.  The results are dependent upon the mass accretion rate; at higher
accretion rates, the fused deuterium is replenished faster, and so the depletion of deuterium occurs
at a higher stellar luminosity (higher mass) (see HCK97 for a more detailed discussion).

It is worth noting that the PS99 calculations of the birthline, 
like those of HCK97 (and Figure 3), assume a 
deuterium to hydrogen ratio of D/H~$= 2.5 \times 10^{-5}$,
whereas recent measurements of this ratio in the local interstellar medium yield
values D/H~$\sim 1.5 \times 10^{-5}$ (Moos \etal 2002).  This would lower birthline
positions in the HR diagram even further.  Stahler's (1988) calculations for
an accretion rate of $\mdot = 10^{-5}$ indicated that the birthline radius 
of a $1 \msun$ star would be reduced from about $5 \rsun$ to about $3 \rsun$ 
by reducing the deuterium abundance from $2.5 \times 10^{-5}$ to $1.3 \times 10^{-5}$.  
Even assuming that the deuterium abundance is not reduced, 
and within the context of the assumption of spherical accretion, 
Stahler (1988) showed significant differences in birthline positions for varying accretion rates. 

As emphasized by HCK97, birthline theory does not preclude stars
from lying above the birthline; the initial conditions of the quasistatic protostellar 
core onto which accretion occurs result from hydrodynamical collapse, 
which is beyond the scope of the calculations (e.g., Bodenheimer \etal 2000).
Deuterium fusion simply prevents the initial core from lying below far below the birthline,
but does not prevent objects from starting above the birthline; such stars will
simply contract rapidly until they reach the birthline and deuterium
fusion begins.  Furthermore, as shown in Figure 3,
if accretion rates vary, stars of the same final mass will have {\em different} birthlines.
Given that T Tauri stars exhibit a wide range of disk accretion rates among objects of
roughly the same age and mass (Gullbring \etal 1998), it is reasonable to expect similar
variations in disk accretion rates for protostellar objects. 
HCK97 also showed that birthlines can differ significantly even at a single accretion rate
for differing initial protostellar core masses and radii.  Finally, accretion onto protostars
is likely to be time-variable (Henriksen, Andr\'e, \& Bontemps 1997) if not episodic;
the effects of such variable accretion have not been quantified but might well add additional
variability to birthline positions.  Thus, it is far more likely
that young stars start their quasi-static contraction after protostellar accretion
on a range of birthlines, or over a birth region, rather than along a single locus in
the HR diagram.

Qualitatively, one would expect that even if deuterium fusion provides a ``floor'' in
the HR diagram, below which very young low-mass stars cannot be found,
this limitation is less likely to apply for higher-mass objects.
Deuterium fusion can only prevent contraction when it provides a source
of energy comparable to that lost from the stellar surface by radiation;
in higher-mass, higher-luminosity stars, deuterium fusion cannot match
the stellar radiative losses for a long period of time.
In addition, once a pre-main sequence star develops a radiative core, mixing of 
freshly-accreted deuterium from the surface, where it is accreted, 
into the core, where it can be fused, 
will be slowed.  Thus, it is qualitatively reasonable
to assume that the birthlines or birth regions of stars with masses
$< 1 \msun$ can be more tightly linked to deuterium fusion
independent of initial conditions and accretion rates, while the birthline for higher-mass
stars will be less controlled by deuterium fusion and therefore be more sensitive
to mass accretion rates and initial conditions.
This is demonstrated by the convergence of the two HCK tracks in
Figure 3 near $0.5 \msun$ and their divergence at $1 \msun$; similar behavior,
though less dramatic, is also shown in the calculations of Stahler (1988).

In summary, I conjecture (as in Paper I) that the discrepancy in age 
distributions between hotter and cooler stars shown in Figure 1, and the similar 
behavior found in other star-forming regions, is due to lower, variable 
birthline positions of the hotter, more massive stars than assumed by PS02.  
This hypothesis should be tested by more
sophisticated evolutionary calculations extending to higher masses, exploring
a variety of initial conditions, and following the effects of time-dependent accretion
rates.  Better empirical constraints on protostellar masses,
radii, and accretion rates also are needed.  However, apart from these
theoretical considerations, in the absence of evidence for such highly-skewed
stellar IMFs in star-forming regions, the PS00/02 results must be
viewed with considerable skepticism.

As an aside, if my ``birth region'' conjecture is correct, the ages of Herbig Ae/Be stars 
have been systematically underestimated.  Moreover, the age estimates would be
much more uncertain in absolute terms than those of the lower-mass T Tauri
stars, due to variations in birthline positions.  More investigation of
this conjecture is needed, as it has important implications for efforts
to understand the evolutionary timescales of circumstellar disks.

\section{Discussion}

\subsection{Extended and ``accelerating'' star formation in time}

The scarcity of significant molecular cloud complexes with extremely low star formation
rates, and the lack of evidence for strongly variable IMFs, with a strong enhancement
of $\gtrsim 1 \msun$ stars relative to $\lesssim 1 \msun$ stars, casts considerable doubt
on the PS00/02 reconstruction of star formation histories.  In the case of Taurus,
removing older stars with strong Li depletion, and setting aside $\gtrsim 1 \msun$
stars because of birthline uncertainties, the evidence for an extended period of
star formation almost completely disappears.  In any event, as emphasized by HBB01,
the vast majority of stars in Taurus have been formed over only a few Myr.  

Among the stars older than 10 Myr, one (LkCa 19) shows no evidence of Li depletion;
two others (HBC 351 and RXJ0423.7) show modest depletion, and another (HN Tau)
is a classical T Tauri star, thus suggesting that all four systems are indeed relatively
young.  However, this does not mean automatically that the rest of the Taurus cloud
was present, forming no stars for $> 10$~Myr, while these stars were formed.
It is possible that the molecular material associated with the formation of
these few stars formed earlier than Taurus, and has already dispersed.
The region south of Taurus, discussed in \S 3, shows that that a few 
relatively young stars unconnected with present molecular material
are to be expected.
Further examples of this possibility are the $\eta$~Cha and TW Hya associations 
(Mamajek, Lawson, \& Feigelson 1999; Kastner \etal 1997; Webb \etal 1999) 
which are $\sim 10$~Myr-old groups of stars not associated with molecular gas, 
at distances from the Sun of $\sim 100$ and 50 pc, respectively.
Understanding the star formation epoch in Taurus and other regions should
rest more heavily on the properties of the great majority of stars rather than
the interpretation of a very small number of outlying objects.

Because star-forming regions have finite lifetimes, all such regions must experience
``acceleration'' in their star formation rates (from zero to a finite value).
The physical significance of this depends upon the precise timescale involved.
I showed in Paper I that observational errors were likely to have a significant
effect on estimates of the star formation rates over modest age ranges.  Moreover,
I showed that estimates of the likely random or quasi-random errors due to distance
uncertainties, binary companions, extinction errors, etc. are generally logarithmic
in nature.  Therefore, linear age binning such as that of Figure 1 has the disadvantage
that the errors are larger on the high age end than on the low age end.  Furthermore,
if the binning is too coarse, there will be a pile-up of stars in the first age bin.
As shown in Figure 2 of Paper I, these effects can cause a spurious impression of
acceleration in the star formation rate toward the present time.  

Paper I showed that the age distribution of most of the Taurus stars is consistent with a modest
age spread of $\sim 4$~Myr; the present results do nothing to change this conclusion.
Inferring the variation of star formation rate with time at a more detailed level
than this requires a better understanding of observational errors than presently
available.

PS02 disagree with these conclusions of Paper I, arguing that if observational
errors were responsible for their results, they would expect as many stars to appear
too young as those which appear too old.  Their argument depends upon the mistaken
assumption that all the errors in question are random.  As shown in the previous
sections, inclusion of non-members and incorrect birthline placement can
result in {\em systematic} errors, overestimating the star formation rates at earlier
times.
 
\subsection{Dynamic vs. quasi-static molecular clouds}

Star formation in Taurus is strongly confined to narrow bands or filaments which
extend over a sizable fraction of the total cloud extent (Hartmann 2002,
and references therein; PS02).  As discussed in Hartmann (2002), these large-scale
filamentary structures suggest that the dominant driving is by large-scale turbulent
motions, as observed in some numerical simulations (e.g., Klessen \&
Burkert 2000, 2001; Mac Low \& Ossenkopf 2000; Ballesteros-Paredes
\& Mac Low 2002).  PS02 argue instead
that the filaments are produced by ``global, quasi-static contraction of the
parent cloud material''.  But it is unclear how the supersonically-turbulent
molecular gas in Taurus can quasi-statically contract, let alone why such quasi-static
contraction would produce filaments.  PS02 argue that the rough match between self-gravity
and turbulent motions supports their notion of quasi-static cloud evolution;
but as shown by Ballesteros-Paredes, Vazquez-Semadeni, and Scalo (1999; also HBB), numerical
simulations frequently show similar cloud gravitational and kinetic energies even though
these clouds are never in virial equilibrium.  Moreover, as discussed in Hartmann (2002),
there is evidence for departures from equilibrium in the dense Taurus filaments,
in that supersonic motions are observed, and the filament line densities may exceed
the gravitational stability limit.

PS02 argue that if the rapid star formation picture were correct, with rapid dissipation
of turbulence and fragmentation in collisionally produced shells or filaments, then
star formation would have to be nearly 100\% efficient.  This argument is only true
if turbulence can be ignored and star formation occurs over an extended period.  
Star formation occurs in regions where the turbulence has dissipated,
which occurs rapidly (Stone, Ostriker, \& Gammie 1998; Mac Low \etal 1998;
Padoan \& Nordlund 1999; Mac Low 1999).
Simulations show intense localized star formation surrounded by extended volumes of
non-star-forming, highly turbulent gas (Padoan \& Nordlund 1999;
Ostriker, Stone, \& Gammie 2001; Klessen \etal 2000; Klessen \& Burkert 2000, 2001;
Heitsch, Mac Low, \& Klessen 2001; Vazquez-Semadeni, Ballesteros-Paredes,
\& Klessen 2002).  If the remaining cloud is then rapidly dispersed by
the energy input of stars, as argued by HBB01, the star formation efficiency will be low.

\section{Conclusions}

The star formation histories of Taurus and other regions derived by PS02 and PS00
imply that most molecular clouds should experience an extended period of very
low star formation, with a strong enhancement in the number of $\gtrsim 1 \msun$ stars relative to
stars with masses $< 1 \msun$.  Current observational evidence contradicts these
predictions, rendering the conclusions of PS02 and PS00 suspect.
I have shown that in the specific case of Taurus, the PS02 inferences of extended
periods of star formation depend almost exclusively on results for stars of masses $\gtrsim
1 \msun$.  Significant foreground contamination in Taurus by older non-members
is present in these more massive stars at ages $\gtrsim 10$~Myr, 
and this is likely to a bigger problem for analyses of more distant regions.  
Even restricting the sample to stars younger than 
about 10 Myr, I have shown that the age distribution derived by PS02 for the 
$\gtrsim 1 \msun$ stars differs significantly from that of the lower-mass stars.
I conjecture that this is due to overestimating the ages of the higher-mass stars
due to birthline effects; variations in birthline positions for the more massive
stars dominate the apparent ages and age spread.  I offer theoretical support for
this conjecture. In addition, there is an observational test; if the PS02 interpretation
and birthlines are correct, there must be a significant fraction of all star-forming
regions with highly variable initial mass functions, in the sense of a vastly reduced
relative population of low mass stars.  Current studies yield no evidence for
this proposition, which favors my birthline/birth region conjecture.

The results for stars with masses $\lesssim 1 \msun$ in Taurus, which constitute the
vast majority of the young stellar population, and are the least affected by non-member
contamination and birthline effects, demonstrate that star formation has commenced
rapidly, as argued in HBB01.  This behavior is consistent with recent numerical simulations
of the interstellar medium and molecular cloud formation and evolution which show
that cloud and star formation is a dynamic process.

I am grateful to Cesar Brice\~no and Kevin Luhman for help with making
figures and providing other information.
This work was supported in part by NASA grant NAG5-9670.

\begin{figure}
\plotone{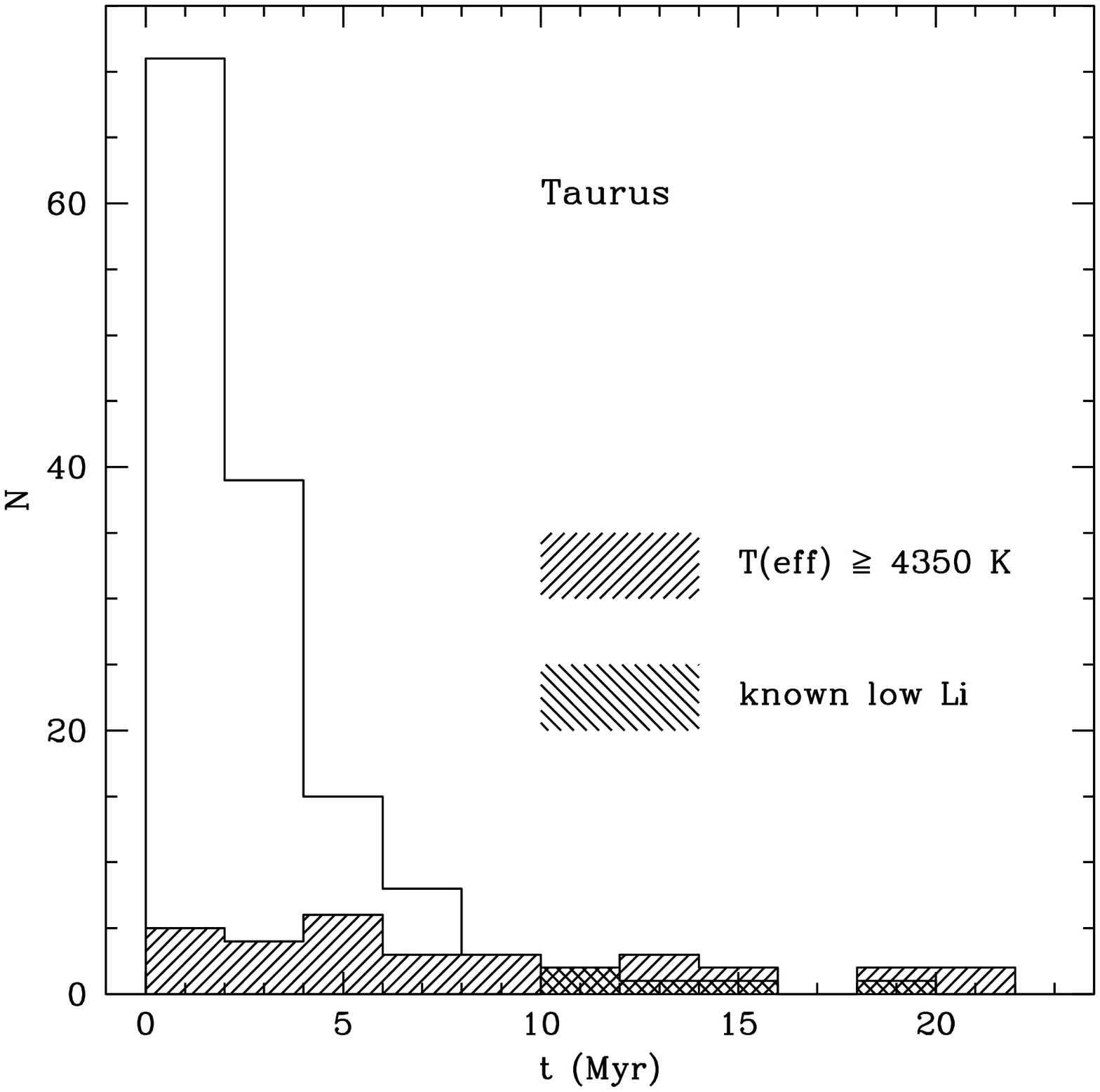}
\caption{Distribution of stellar ages in Taurus from PS02, with
hatching indicating
hotter stars (effective temperatures $\geq 4350$~K) and stars
known to have Li abundances lower than that of main
sequence stars in the young open clusters IC 2391 and IC 2602 (see text)} 
\end{figure}

\begin{figure}
\plotone{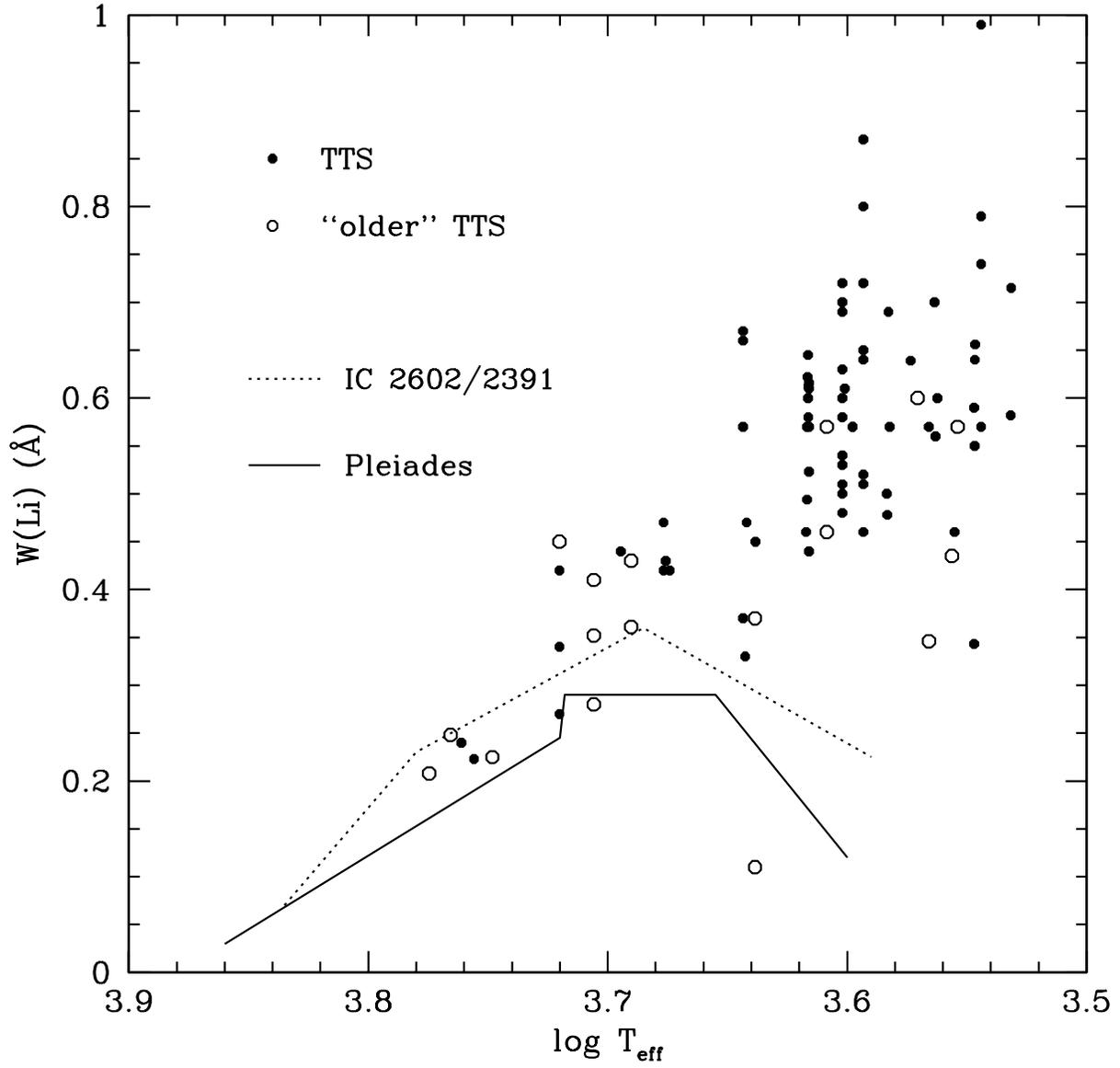}
\caption{Li I 6707 \AA\ equivalent widths
for T Tauri stars (filled dots) and ``older'' Taurus stars from Table 1
(open circles).  The upper envelope of observed equivalent widths
for the Pleiades (solid segmented curve) and the IC 2391/2602 clusters
(dotted segmented curve) are shown for reference} 
\end{figure}

\begin{figure}
\plotone{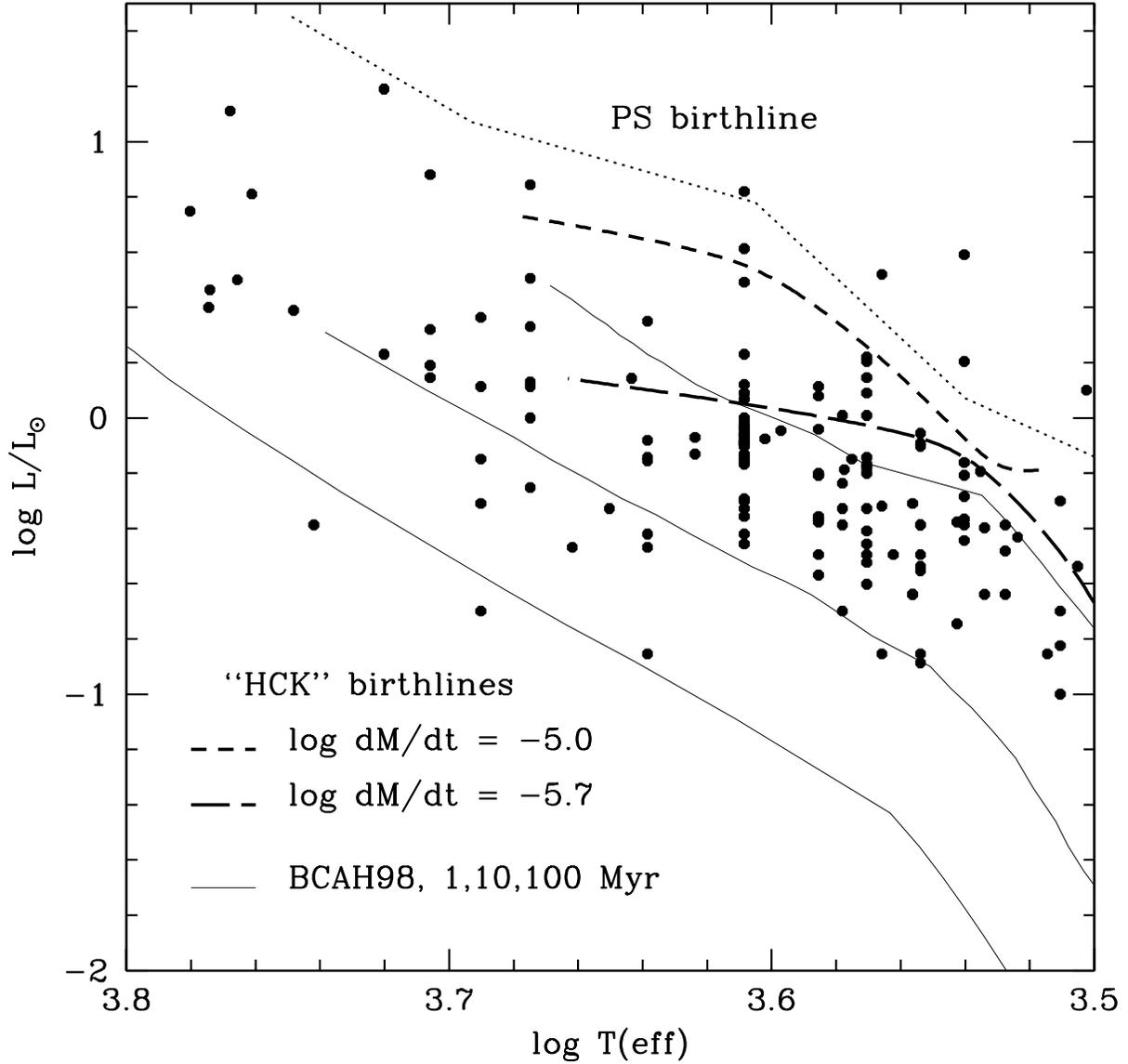}
\caption{HR diagram for Taurus objects from PS02, with 1, 10, and 100
Myr isochrones from Baraffe \etal (1998).  Birthlines are shown 
for PS02 (dotted curve) and for calculations of HCK at two differing
accretion rates, $10^{-5} \msunyr$ and $2 \times 10^{-6} \msunyr$,
modified for fits to the Baraffe \etal (1998) tracks (see text)}
\end{figure}

\newpage

\begin{deluxetable}{rlccrrrr}
\tablecolumns{8}
\tablewidth{0pc}
\tablecaption{Taurus stars from PS02, ages $>$ 6 Myr}
\tablehead{
\colhead{N} & \colhead{Name}\tablenotemark{a} & \colhead{Type}\tablenotemark{b} &
\colhead{T(eff)} & \colhead{$L/\lsun$} & \colhead{Age (Myr)} &
\colhead{W(Li) (\AA)} & \colhead{Ref}\tablenotemark{d}}
\startdata
1  & HBC 351     & W & 4350  & 0.38 &12.1 & 0.37  &B91  \\
2  & RXJ0405.3   & W & 5080  & 1.55 & 9.6 & 0.352 & W  \\
3  & RXJ0405.7   & W & 5830  & 3.16 &10.8 & 0.248 & W  \\
4  & RXJ0406.7   & W & 5950  & 2.51 &16.0 & 0.208 & W  \\
5  & RXJ0412.8   & W & 3600  & 0.23 & 6.5 & 0.435 & W  \\
6  & V410 A24    & x & 5945  & 2.91 &12.6 & \ldots   &  \\
7  & V410 A20    & x & 4730  & 0.56 &20.7 & \ldots   &  \\
8  & HBC 376     & W & 4060  & 0.38 & 6.0 & 0.46  & M94  \\
9  & RXJ0423.7   & W & 4900  & 0.71 &20.0 & 0.361 & W  \\
10 & RXJ0424.8   & W & 5080  & 2.09 & 6.6 & 0.410 & W  \\
11 & HBC 388     & C & 5080  & 1.40 &12.3 & 0.28  & B91  \\
12 & UX Tau A    & W & 4900  & 1.30 & 8.0 & 0.43  & MMR   \\
13 & UX Tau B    & W & 3720  & 0.35 & 7.5 & 0.60  & MMR  \\
14 & HBC 392     & W & 4350  & 0.34 &18.3 & 0.11  & M94  \\
15 & L1551-55    & W & 4060  & 0.35 & 6.8 & \ldots    &  \\
16 & HN Tau      & C & 4350  & 0.70 &22.0 & \ldots    &  \\
17 & HBC 412     & W & 3580  & 0.41 &20.4(4)\tablenotemark{c}&0.57 &  FW  \\
18 & HP Tau/G2   & C & 6030  & 5.60 & 8.4  &\ldots    &  \\
19 & RXJ0438.2   & W & 3680  & 0.14 & 8.0  &0.346 &  W  \\
20 & RXJ0441.8   & W & 5600  & 2.45 &11.   &0.225 &  W  \\
21 & DS Tau A    & C & 4350  & 0.72 & 7.8  & \ldots    &  \\
22 & GM Aur      & W & 4730  & 1.00 & 9.6  & \ldots    &  \\
23 & LkCa 19     & W & 5250  & 1.70 &14.4  & 0.45   &B91  \\
24 & V836 Tau    & C & 4060  & 0.51 & 7.9  & 0.57   &S  \\
\enddata
\tablenotetext{a}{HBC numbers from the catalog of Herbig
\& Bell (1998)} 
\tablenotetext{b}{W $=$ WTTS, C $=$ CTTS}
\tablenotetext{c}{The age of HBC 412 is estimated
at 4 Myr (see text)}
\tablenotetext{d}{References: FW $=$ Walter \etal 1988;
B91 $=$ Basri \etal 1991; M94 $=$ Martin \etal 1994; MMR $=$ 
Magazzu \etal 1991; S $=$ Strom \etal 1989; W $=$ Wichmann \etal 2000.}
\end{deluxetable}

\end{document}